\documentclass[twoside]{article}
\usepackage{fleqn,espcrc2}
\usepackage{graphicx}

\newcommand{\AmS}{{\protect\the\textfont2
  A\kern-.1667em\lower.5ex\hbox{M}\kern-.125emS}}
\usepackage{graphicx}%style pour inserer figures en eps
\usepackage{here}%pour inserer figures et tables a emplacement 
\def\beq{\begin{equation}}
\def\eeq{\end{equation}}
\def\bea{\begin{eqnarray}}
\def\eea{\end{eqnarray}}
\def\bq{\begin{quote}}
\def\eq{\end{quote}}

\def\bear{\begin{array}}
\def\ear{\end{array}}

\def\ga{\left(}
\def\dr{\right)}

\def\rar{\rightarrow}

\def\la{\langle}
\def\ra{\rangle}
\def\nin{\noindent}
\def\ba{\begin{array}}
\def\ea{\end{array}}

\def\b{\bullet}

\def\gam5{\gamma_5}

\title{\bf{Scalar mesons in QCD and tests of the gluon content of the $\sigma$}
\thanks{Mini-review given at the QCD 02 9th International  High-Energy Physics Conference in Quantum ChromoDynamics (Montpellier
2-9th July 2002).}}
\author{ Stephan Narison\address{
Laboratoire de Physique Math\'ematique,
Universit\'e de Montpellier II
Place Eug\`ene Bataillon,
34095 - Montpellier Cedex 05, France. 
\\ E-mail:
qcd@lpm.univ-montp2.fr}}
\begin{document}
\pagestyle{empty}
\begin{abstract}
\noindent
We summarize the different features of the scalar mesons from QCD spectral sum rule 
analyses of the two- and three-point  functions. The results do not
favour the $\bar uu+\bar dd$ interpretation of the broad and low mass $\sigma 
(0.6)$, and the $\bar us$ resonance nature of the eventually observed $\kappa(0.9)$ meson. We also discuss some OZI-violating
and classic semileptonic and radiative decay processes which can reveal in a 
 model-independent way the eventual gluon component $\sigma_B$ of the $\sigma$. 
In a meson-gluonium mixing scenario, one also expects  an observation of the $K\bar K$ final states from the $\sigma_B$ which may
compete (if phase space allowed) with the one from a low mass
$\bar ss$ state assumed in the literature to be the SU(3) partner of the $\sigma (0.6)$ if this latter is a
$\bar uu+\bar dd$ state.\vspace*{2mm}
\noindent
\end{abstract}
\maketitle
\section{Introduction}
\nin
The nature of scalar mesons is an intriguing problem in QCD. Experimentally,
there are well established scalar mesons with isospin $I=1$, the $a_0(980)$, with isospin $I=1/2$
$K^*_0(1410)$ meson,  and with isospin $I=0$, the $f_0$-mesons at 980, 1370 and 1500 MeV \cite{PDG}. Besides these
resonances there are different indications \cite{MONTANET,KYOTO} for a
low lying scalar isoscalar state, the famous $\sigma$. 
The isoscalar scalar states are especially interesting in the
framework of QCD since, in this $U(1)_V$ channel, their interpolating operator is the trace of the
energy-momentum tensor:
\beq
\theta_\mu^\mu=\frac{1}{4}\beta(\alpha_s) G^2+\sum_i [ 1+\gamma_m(\alpha_s)]
m_i\bar\psi_i\psi_i~,
\eeq
where $G^a_{\mu\nu}$ is the gluon field strengths, $\psi_i$ is the quark field; $\beta(\alpha_s)$ and
$\gamma_m(\alpha_s)$ are respectively the QCD $\beta$-function and quark mass-anomalous dimension. In the chiral
limit
$m_i=0$,
$\theta_\mu^\mu$ is dominated by its gluon component $\theta_g$, like is the case of the $\eta'$ for the
$U(1)_A$ axial-anomaly, explaining why the $\eta'$-mass does not vanish like other Goldstone bosons for $m_i=0$. In
this sense, it is natural to expect that these $I=0$ scalar states are glueballs/gluonia or have at least a strong
glue component in their wave function. QCD spectral sum sum rules (QSSR) are an important analytical tool of
nonperturbative QCD and especially well suited to address the question of the quark-gluon mixing since the
principal nonperturbative ingredients are the quark condensate, the gluon condensate and the mixed quark-gluon
condensate. 
In this talk, we summarize some essential features of QSSR analyses
and discuss different production processes 
for obtaining information on the gluon content of the scalar mesons. 
%%%%%%%%%%%%%%%%%%%%%%%%%%%%%%%%%%%%%%%%%%
\section{Unmixed scalar mesons from QSSR}
\nin
$\b$ Masses and couplings of unmixed scalar $\bar qq$ mesons and gluonia have been extensively studied in the past using
QSSR within the standard Operator Product Expansion (OPE) of the diagonal two-point correlator \cite{SNG,SNB}:
\beq
\psi(q^2)=i\int d^4x e^{iqx}\la 0|{\cal T}J(x)J(0)^\dagger |0\ra~,
\eeq
associated to the quark
%$$\pr^\mu A^{ij}_\mu=(m_i-m_j):\bar \psi_i\psi_j:~J_2=m:(\bar uu+\bar dd):~{\rm or}~J_3=2m_s:\bar ss: $$
or/and the gluonic currents
%$$J_g\equiv 4\theta_g=\beta(\alpha_s) G^2~.$$
It has been emphasized that the mass of the isoscalar scalar $S_2\equiv \bar uu+\bar dd$ meson is about 1 GeV, in agreement with
the one of the observed $a_0(980)$, as expected from the good realization of the $SU(2)$ symmetry implying a degeneracy between the
$a_0$ and $S_2$ states, while its width into $\pi\pi$ is about 100 MeV \cite{SNB,SNG}. On the other, the mass of the mesons
containing a strange quark is above 1 GeV due to $SU(3)$ breaking, where the same mechanism explains successfully the well-known
$\phi$--$\rho$ and
$K^*$--$\rho$ mass splittings. In recent analysis~\cite{Dosch:2002rh,DSN02}, it has been shown that instanton within the instanton
liquid model \cite{SHURYAK} as well as new $1/q^2$ \cite{DSN02} induced by a tachyonic gluon mass \cite{CNZ} from the
linear term of the short distance part of the QCD potential \footnote{A linear term of the potential at all distances has been
recently proposed by 't Hooft \cite{THOOFT} as a possible way to solve the confinement problem.}, affect only slightly these mass
predictions and do not allow to decrease these values in a stable way.\\
$\b$ In the gluonium channel, using a subtracted sum rule sensitive to $\psi_G(0)\simeq -16{\beta_1/ \pi}\la\alpha_s
G^2\ra$, (where $\beta_1=-1/2(11-2n/3)$ and $\la \alpha_s G^2\ra\simeq 0.07$ GeV$^4$ \cite{SNG2}) and the unsubtracted sum rule, it was
found
\cite{VENEZIA,SNB,SNG} that one needs two resonances for consistently saturating the two sum rules, where the lowest mass
gluonium
$\sigma_B$ should be below 1 GeV. A low energy theorem obeyed by the vertex $\la\pi|\theta^\mu_\mu|\pi\ra$ also shows that the
$\sigma_B$ can be very wide with a $\pi^+\pi^-$ width of about $(0.2-0.8)$ GeV corresponding to a mass of $(0.7-1)$ GeV, while a
$\sigma_B$ having a mass less than 0.6 GeV  has a weaker coupling because $g_{\sigma\pi\pi}\sim M^2_\sigma$ and thus cannot be broad.
This result shows {\it a huge violation of the OZI rule} like in the case of the $\eta'$-channel \cite{SHORE}.
\\ 
$\b$ From the previous results, one can already conclude that {\it the observed
$\sigma(0.6)$ cannot be a pure $\bar qq$ state, but contains most probably a large gluon component in its wave function \footnote{Similar
conclusions using different approaches have been reached in \cite{OCHS,MONTANET}.},
while the
$\kappa(0.9)$ meson cannot also be identified with an usual $\bar us$ scalar resonace which mass is predicted to be about the
$K^*_0(1.3-1.4)$}. \\
$\b$ As a consequence, a quarkonium-gluonium mixing (decay mixing \footnote{This has to be contrasted with the small mass-mixing coming
from the off-diagonal two-point function \cite{PAK}.}) scheme has been proposed in the
$I=0$ scalar sector
\cite{BN}, for explaining the observed spectrum and widths of the possibly wide $\sigma (< 1$ GeV) and the narrow $f_0(0.98)$. The data are well
fitted when the mixing angle is maximal:\beq
|\theta_S|\approx 40^0~,
\eeq
indicating that the $\sigma$ and $f_0$ have equal numbers of quark and gluon in each of their wave functions. This mixing
scenario also implies a strong coupling of the $f_0$ to $\bar KK$ (without appealing to a four-quark state model) with a strength
\cite{BN}:
\beq
g_{f_0K^+K^-}=2g_{f_0\pi^+\pi^-}~,
\eeq
a property confirmed by the data. The physical on-shell $f_0$ is narrow $(< 134$ MeV) due to a destructive mixing,
whilst the $\sigma(.7- 1)$ can be $(0.4- 0.8)$ GeV wide (constructive mixing). Compared to the
four-quark states and/or $\bar KK$ molecules models (see e.g. \cite{BLACK}), this quarkonium-gluonium mixing scenario includes all QCD
dynamics based on the properties of the scale anomaly $\theta_\mu^\mu$, which comes from QCD first principles. We propose some further
tests of this scenario from semileptonic and radiative decay processes in the following \footnote{We shall not discuss hadronic
productions which are less clean than the former two due to eventual rescattering effects.}.
%%%%%%%%%%%%%%%%%%%%%%%%%%%%%%%%%%%%%%%
\section{Tests from $D_{(s)}$ semileptonic decays}
%%%%%%%%%%%%%%%%%%%%%%%%%%%%%%%%%%%%%%%
\subsection*{$S_2(\bar uu+\bar dd)$ meson productions}
%%%%%%%%%%%%%%%%%%%%%%%%%%%%%%%%%%%%%%%%%%%%%%%%%%%%%%
\nin
$\b$ If the scalar mesons were simple $\bar qq$ states, the semileptonic decay width
could be calculated quite reliably using QSSR, where the relevant diagram is a quark loop triangle.
This analysis has
been done with a good success for the semileptonic decays of the $D$ and $D_s$
into pseudoscalar and vector mesons \cite{SNB}. For the production of a pseudoscalar or
scalar $\bar qq$ states several groups \cite{Dosch:2002rh} predict all form factors to be:
%\beq
$f_+(0)\approx 0.5~,$
%\eeq This
yielding a decay rate:
\beq
\Gamma(D\rar S_2 l\nu)=(8\pm 3)10^{-16} ~{\rm GeV}~,
\eeq
for $M_{S_2}\simeq 600$ MeV.\\
$\b$ However, because of the enigmatic nature of the  $\sigma$,  one has also considered,  in ~\cite{Dosch:2002rh}, the
case that the quark-antiquark current does not couple to
a resonance but rather to an uncorrelated quark-antiquark pair. In that case
the decay rate is reduced by a factor 2, but, in the spectral distribution,
there is a broad bump visible with a maximum near the presumed
$\sigma$ mass of 600 MeV. Unfortunately, even in high stastistics
experiments, the estimated decay rates of the $D$-meson are at the edge of observation since the
decays into an isoscalar are CKM-suppressed due to the $c$-$u$ transition at
the weak vertex.
%%%%%%%%%%%%%%%%%%%%%%%%%%%%%%%%%%%%%%%%%%%%%%%%%%%%%%%%%%
\subsection*{Scalar gluonium and/or $\bar ss$ productions}
\nin
The evaluation of 
diagrams for a semileptonic decay into a gluonium state 
is, unfortunately, more involved than in the $\bar qq$ case.
Therefore, we can give only semi-qualitative results which, however, are
model independent.\\
$\b$ The only way to obtain a non-CKM suppressed isoscalar is to look at the
semileptonic decay of the $D_s$-meson, where the light
quark is a strange one and an isoscalar $s \bar s$ or/and gluonium state can be
formed. \\
$\b$ If the $\bar ss$ is relatively light ($<1$ GeV), which might be the natural partner of the $\bar
uu+\bar dd$ often interpreted  to be a $\sigma (0.6)$ in the literature, then, one should produce a
$\bar KK$ pair through the isoscalar $\bar ss$ state. The QSSR prediction for this process is under quite good
control \cite{SNB,Dosch:2002rh}. The
non-observation of this process will disfavour the
$\bar qq$ interpretation of the $\sigma$ meson. \\
$\b$ If a gluonium state is formed it will decay with even strength into $\pi\pi$ and a $K\bar K$ pairs. Therefore a gluonium
formation in semileptonic $D_s$ decays should result in the decay patterns: 
\beq
D_s\rar \sigma_B\ell \nu\rar \pi \pi \ell \nu ~~~~~ D_s\rar\sigma_B\ell \nu\to K \bar K  \ell \nu~,
\eeq
with about the same rate up to phase space factors.
{\it The observation of the semileptonic $\pi\pi$ decay of the $D_s$ would be a unique
sign for glueball formation.}\\
$\b$ A semi-qualitative estimate of the above rates can be obtained by working in the large heavy quark mass limit $M_c$. Using,
e.g., the result in ~\cite{Dosch:2002rh}, the one for light $\bar qq$ quarkonium production behaves as:
\beq
\Gamma[D_s\rar S_q(\bar qq)~l\nu]\sim |V_{cq}|^2G^2_F M^5_c |f_+(0)|^2~.
\eeq
$\b$ For the $\sigma_B(gg)$ production, we study the $1/M_c$ behaviour of the $WW gg$ box diagram,
where it is easy to find that the dominant (in $1/M_c$) contribution comes from the one involving one charmed propagator. Therefore the
production amplitude can be described by the Euler-Heisenberg effective interaction :
\beq
{\cal L}_{eff}\sim \frac{g_W\alpha_s}{p^2M_c^2}F_{\mu\nu} F^{\mu\nu}G_{\alpha\beta}G^{\alpha\beta}+ {\rm
perm.}+\cdots
\eeq
where $\cdots$ are h.o in $1/M_c$, $g_W$ is the electroweak coupling and $p^2\simeq M^2_{\sigma}$ is the typical virtual low scale
entering into the box diagram. Using dispersion techniques similar to the one used for
$J/\psi\rar \sigma_B \gamma$ processses \cite{NSVZ,VENEZIA,SNB}, one obtains, assuming a $D_s$ and
$\sigma_B$-dominances:
\beq
\Gamma[D_s\rar\sigma_B(gg)~l\nu]\sim |V_{cs}|^2G^2_F  \frac{|\la 0|\alpha_s G^2|\sigma_B\ra|^2}{M_cM_\sigma^4}.
\eeq
The matrix element $\la 0|\alpha_s G^2|\sigma_B\ra $ is by definition proportional to $f_\sigma M^2_\sigma$, where $f_\sigma$ is
hopefully known from two-point function QSSR analysis \cite{VENEZIA,SNG,SNB}. Using $f_\sigma\approx 0.8$ GeV, one
then deduces:
\beq
{\Gamma[D_s\rar\sigma_B(gg)~l\nu]\over \Gamma[D_s\rar S_q(\bar qq)~l\nu]}
\sim {1\over |f_+(0)|^2}\ga{f_\sigma \over M_c}\dr^2,
\eeq
which is $ {\cal O}(1)$. This qualitative result indicates that {\it the gluonium production rate can be of the same order as the
$\bar qq$ one contrary to the na\"\i ve perturbative expectation ($\alpha_s^2$ suppression), which is a
consequence of the OZI-rule violation of the $\sigma_B$ decay.}\\
$\b$ However, it also indicates that, due to the (almost) universal coupling of the $\sigma_B$ to Goldstone boson pairs, one also
expects a production of the $K\bar K$ pairs, which can compete with the one from $\bar ss$ quarkonium state, and again renders more
difficult the identification of the such $\bar ss$ state if allowed by phase space.
%%%%%%%%%%%%%%%%%%%%%%%%%%%%%%%%%%%%%%%%%%%%%%%%%%%%%%%%%%%%%%%%%%%
\section{Tests from $J/\psi$ and $\phi$ radiative decays}
\nin
Radiative decays of vector mesons have been often proposed to be the classical gluonia production processes. In \cite{SNG,VENEZIA},
one expects the rate:
\beq
B[J/\psi\rar \gamma\sigma_B] B[\sigma_B\rar all]\approx (4-6)\times 10^{-4},
\eeq
which is far below the upper rate of production for $B(J/\psi\rar \gamma\pi\pi)$ of about $10^{-2}$ allowed by BES \cite{MONTANET}.
Analogous rate is expected for the $\sigma'(1.3)$(radial excitation of the $\sigma$) and $G(1.5)$ productions.\\
$\b$ Extending the analysis of the $J/\psi$ into the one of the $\phi$ by replacing the charm quark constituent loop by the strange
quark, it is easy to find \cite{SNUNPUB}:
\beq
Br[\phi\rar\gamma f_0(980)]\approx 1.3\times 10^{-4}~,
\eeq
which despite the crude approximation (use of the $1/m_s$ constituent mass expansion) leads to a surprising satisfactory
result compared with the averaged data
 $(1.08\pm 0.07\pm 0.06)\times 10^{-4}$ from \cite{CMD}.\\
$\b$ For the $a_0$, one also expects that the
L$\sigma$M + ChPT ($SU(3)$ symmetry) can give a reliable prediction of the production rate (see e.g.
\cite{ESCRIBANO}) due to its isovector nature contrary to the case of the isoscalar mesons affected by their gluon content
($U(1)_V$ symmetry). In this case, one expects the decay chain process through kaon loops:
\beq
\phi\rar \gamma\bar KK\rar \gamma a_0\rar\gamma\eta\pi~,
\eeq
where the coupling of the $a_0$ to $\bar KK$ can be obtained from e.g. the $SU(3)$ relation in \cite{BN}.
%%%%%%%%%%%%%%%%%%%%%%
\section{Conclusions}
\nin
We have reviewed the different features of scalar mesons from QCD spectral sum rules
%, which do not favour the $\bar uu+\bar dd$
%interpretation of the broad and low mass $\sigma (0.6)$ and $\kappa(0.9)$. 
and the different OZI-violating and
classic gluonia semileptonic and radiative decay processes for testing the eventual gluon component of the
$\sigma$ meson. We wish that some progresses on this field will be accomplished in the near future, as
after about a 1/4 century study, we still remain with more questions than answers on the true nature of scalar mesons.  
%%%%%%%%%%%%%%%%%%%%%%%%%%%
\section*{Acknowledgements} 
\nin
This talk has been completed  at the Max-Planck-Institute f\"ur Physics of Munich. It is a pleasure
to thank Valya Zakharov for his invitation and for discussions. 

%%%%%%%%%%%%%%%%%%%%%%%%%%%%%%%%%%%

\end{document}